\documentclass[preprint,superscriptaddress]{revtex4-2}

\usepackage{graphicx}   
\usepackage{subcaption}     
\usepackage{float}       
\usepackage{dcolumn,multirow}
\usepackage{bm}
\usepackage{amsmath,amssymb}
\usepackage{hyperref}
\usepackage{xcolor, soul}
\usepackage{adjustbox} 
\usepackage{booktabs}   
\usepackage{siunitx}    
\usepackage{makecell}
\graphicspath{{Figures/}}
\begin{document}

\title{\textbf{Intertwined Hyperferroelectricity, Tunable Multiple Topological Phases and Giant Rashba Effect in Wurtzite LiZnAs}}

\author{Saurav Patel}
\affiliation{Department of Physics, Faculty of Science, The Maharaja Sayajirao University of Baroda, Vadodara-390002, Gujarat, India.}

\author{Paras Patel}
\affiliation{Department of Physics, Faculty of Science, The Maharaja Sayajirao University of Baroda, Vadodara-390002, Gujarat, India.}

\author{Shaohui Qiu}
\affiliation{Department of Chemistry and Physics, Southern Utah University, Cedar City, Utah 84720, USA.}

\author{Prafulla K. Jha}
\email{prafullaj@yahoo.com}
\affiliation{Department of Physics, Faculty of Science, The Maharaja Sayajirao University of Baroda, Vadodara-390002, Gujarat, India.}

\begin{abstract}
	Composite quantum compounds (CQCs) offer a fertile ground for uncovering the complex interrelations between seemingly distinct phenomena in condensed matter physics for advanced non-volatile and spintronics applications. Beyond topological superconductors and axion insulators, the idea of intertwined  Hyperferroelectricity (HyFE), tunable multiple topological phases and Rashba spin-splitting with reversible spin textures represents the local, global and symmetry-driven characteristics of quantum materials, respectively, offering unique pathways for enhanced functionalities. We unveiled a unified framework to achieve this synergy through the presence of crystalline symmetries and spin-orbit coupling (SOC) in LiZnAs compound using \textit{first-principles} calculations. HyFE materials exhibits rare ability to maintain spontaneous polarization under open-circuit boundary conditions, even with the existence of depolarization field while Rashba effect exhibits paradigmatic spin texture in momentum space with tangential vector field. The presence of unstable $A_{2u}(LO)$ mode leads to the free energy minimum with significant well depth and polarization of -66 meV and $P_\mathrm{HyFE} = 0.282~\mathrm{C/m^2}$, respectively indicating stable HyFE. This is highest among intrinsically explored materials. The robust HyFE stem from \textit{mode-specific effective charges} (MEC) and larger high-frequency dielectric constants. This study also addresses the subtle question of whether the critical point of topological phase transition shifts in response to the drastically different Rashba spin-splitting values obtained from VASP and WIEN2k codes. Moreover, the biaxial strain (BAS) induced Weyl semimetal (at 3.4\% BAS) and topological insulating phase (after 3.4\% BAS) is observed with giant Rashba coefficient of 5.91 eV \text{\AA} and 2.42 eV \text{\AA}, respectively. Furthermore, switching of bulk polarization leads to spin texture reversal, providing a robust mechanism to leverage spin degrees of freedom in these Hyperferroelectric Rashba topological materials. This research initiative unlock pathways with more reliable theoretical approaches including rigorous free energy approach along with MEC analysis in HyFE and use of full-potential methods for accurate Rashba spin-splitting.
	
\end{abstract}

\maketitle

\section{\label{1}Introduction}
\vspace{-10pt}
As Moore's law approaches its fundamental limits, the search for transformative approaches in nanoelectronics has intensified. Consequently, there has been surge rise to unveil the potentials of topologically protected fermionic excitations, spin degrees of freedom along with electric polarization in single material to engineer the next wave of nanoelectronics\cite{Hasan2010,Das2004,DiSante2016}. The coexistence of multiple quantum phenomena within a single material offers a powerful platform to explore the underlying physical mechanisms and their mutual interplay to leverage enhanced functionalities\cite{Gupta2023}. The materials exhibiting such multiple quantum behavior are known as composite quantum compounds (CQCs)\cite{Li2019MnBi2Te4}. Recently, topological supercondutors\cite{Hasan2010}, topological axion insulators\cite{Liuaxion2020} and topological insulator with Rashba spin-splitting (RSS)\cite{Mondal2021} emerge as potential CQCs. However, they are limited to coexistence of only two quantum phenomena and the search for intertwined multiple quantum properties within single material remains an active and pressing frontier of research.

In recent years, crystalline symmetry and relativistic spin-orbit coupling (SOC) plays a pivotal role in condensed matter physics by enabling diverse phenomena such as topological quantum phases and relativistic spin-splitting mechanisms akin to Rashba and Dresselhaus effects\cite{DiSante2016,Yoshimi2025,Manchon2015}. In particular, topological insulators (TIs) exhibit protected surface states with massless spin-polarized Dirac fermions, whereas asymmetry in noncentrosymmetric systems leads to spin-splitting in bands of massive fermions. The spin-momentum locking in both cases offers a path toward complimentary electric control of spins along with dissipation-less spin current, making them excellent for spintronic applications\cite{Saurav2025}. In general, the topological phase transitions (TPT) among topological insulators (TIs), gapless Dirac/Weyl semimetals, and trivial insulators are governed by symmetry breaking and band inversion, as described in the phase diagram by Murakami \textit{et al}.\cite{Murakami2008}. This leads to the revived interest in time-honored SnTe compound with broken inversion symmetry induced coexistence of Rashba and non-trivial topology\cite{Hsieh2012}. Interestingly,  the class of SnTe materials exhibit intrinsic ferroelectric (FE) instability owing to their rock-salt structural geometry\cite{Plekhanov2014} which facilitates the Weyl semimetal phase near TPT indicating exotic nonlinear Hall effect driven by Berry curvature dipole and enhanced anomalous Hall conductivity\cite{Yang2018, Zhang2022}. Moreover, the intrinsic coupling between FE polarization and spin texture allows for reversible switching, a hallmark feature for promising non-volatile spintronic applications\cite{DiSante2016}. Consequently, FE Rashba semiconductors have emerged, with GeTe recognized as a representative prototype\cite{DiSante2013}. At surfaces and interfaces lacking inversion symmetry, spin–orbit coupling gives rise to the celebrated Rashba effect, which splits otherwise degenerate electronic bands. The strong Rashba effect arise in compounds with narrow band gaps and heavy constituents; enabling precise control of electron spins and facilitating efficient spin transport, relaxation, and detection processes. With the growing interest, several polar materials are investigated to understand the intercorrelation between topology, Rashba spin-splitting (RSS) and ferroelectricity\cite{Gupta2023,Tian2024,Yang2024}. In particular, BiTeI exhibits giant Rashba splitting and topological behavior under large pressure. The class of LiGaGe-type compounds predicted by Bennett \textit{et al.}, \cite{Bennett2012} have attracted considerable attention for exploration of Rashba, Dirac and Weyl fermions within single materials owing to its moderate band gaps and heavy constituents\cite{Narayan2015,Mondal2021}. Notably, our previous investigation on Group I-IV-V based compounds for potential CQCs conclude highest reported Rashba coefficient and physical origin behind the necessity of full-potential density functional theory (DFT) framework for accurate prediction of RSS along with intrinsic ferroelectricity and topological phases \cite{Saurav2025}. It is therefore pertinent to ask \textit{if the quantum critical point of a TPTs shifts in response to differing Rashba effect magnitudes predicted by pseudopotential and full-potential methods}, given that TPTs are inherently governed by the evolution of the electronic bands under functionalization.

A FE phase transition is generally triggered by an unstable zone-center transverse-optic (TO) mode associated with symmetry-lowering polar displacements from a non-polar reference phase\cite{Vanderbiltperovskite1994}. In FE slab with insulating surfaces, spontaneous polarization perpendicular to the surface is suppressed under open-circuit boundary conditions (OCBC) due to the high energetic cost of the depolarization field, which outweighs the distortion energy gain. In practical applications, metallic electrodes are placed at the surfaces to screen the depolarization field\cite{Rappe2005}. The absence of ferroelectricity under OCBC restricts the miniaturization of FE devices and hinders its practical applications. In 2014, Garrity \textit{et al.},\cite{Garrity2014} introduced a new class of materials, termed as hyperferroelectrics (HyFE) with persistent out of plane polarization even under OCBC. This inherent polarization stability facilitates reduced coercive fields for polarization switching\cite{Qiu2021},enhanced storage density in FE memories \cite{Qiu2023}, and the emergence of polarization tristability and negative piezoelectric coefficients \cite{Liu2017}, underscoring both their fundamental interest and technological potential. The confluence of exotic HyFE properties with topological surface states (TSS) and RSS can pave the way for new avenues of interplay between these distinct physical phenomena. In particular, the occurrence of soft longitudinal-optic (LO) phonon modes at the zone-center are widely recognized as the hallmark of HyFEs. Accordingly, hexagonal ABC-type semiconductors\cite{Garrity2014} and LiBO$_{3}$ (B = V, Nb, Ta, Os)\cite{Li2016} have been proposed as promising HyFE candidates. The possible emergence of hyperferroelectricity has been attributed to reduced LO-TO splitting, typically arising from small Born effective charges\cite{Garrity2014}. Nevertheless, Li \textit{et al}.,\cite{Li2016} demonstrated that such a condition is not strictly required, citing the presence of HyFE nature in LiBO$_{3}$ type structures where Born charges are significantly large. Other potential mechanisms include the depth of internal energy well, low spontaneous polarization\cite{Fu2014}, dominant short-range interactions\cite{Khedidji2021} and the effect of meta-screening\cite{Zhao2018}. Moreover, recent findings from the co-authors of this work shows that LiNbO$_{3}$ fails to exhibit a non-zero polarization at the free energy minimum under OCBC, thereby violating the HyFE definition despite satisfying LO softening criteria\cite{Qiu2021}. These findings imply that soft LO modes are required, yet insufficient on their own to validate HyFEs. Altogether, this highlight the complexities and unresolved challenges in identifying and validating HyFE materials. While the field is still in a beginning  stage, recent methodological advances and deeper theoretical insights proposed by one of the co-authors of this work\cite{Qiu2021,Qiu2023,Adhikari2019} have revived interest in uncovering the physical origin and validation of HyFE nature in candidate materials.

\begin{figure}[H] 
	\centering
	\includegraphics[width=0.70\linewidth]{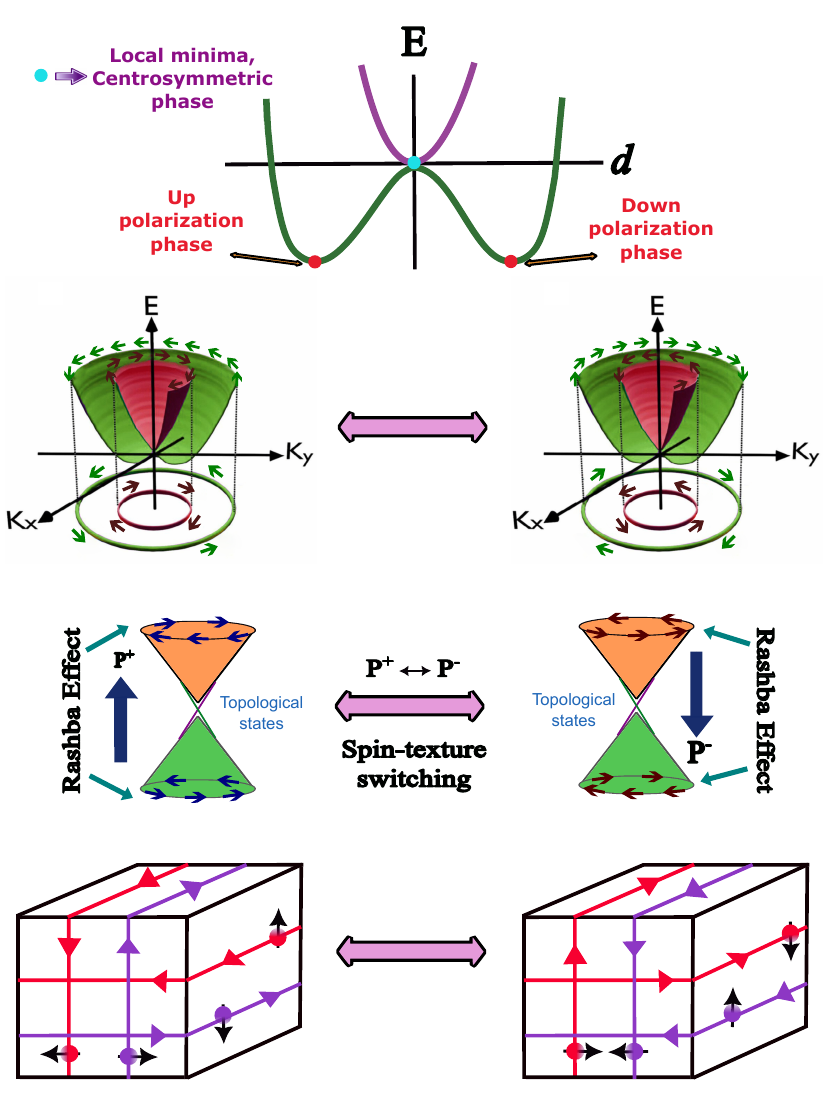}
	\caption{Schematic illustrating the intercorelation among ferroelectric distortion, Rashba spin-splitting, intertwined Rashba effect and topological states and switching mechanism of topological insulators, respectively.}
	\label{Fig:schematic}
\end{figure}

The aforementioned fields are currently at the forefront of cutting-edge research in quantum materials and multifunctional applications. This article aims to discuss unified approach for the exploration and choice of methods for the accurate prediction of novel CQCs with HyFE, RSS along with multiple topological phases within single material platform. Altogether, such CQCs offer a compelling platform where intrinsic HyFE persists polarization under OCBC along with complimentary screening of depolarization field from TSS\cite{Shi2016}. The schematic illustrating energy profile for non-polar and polar phases as a function of distortions as well as their interplay with Rashba topological insulators is represented in Fig.~\ref{Fig:schematic}. It is also evident that switching the polarization of materials leads to spin texture reversal among two minimum configurations. Moreover, one can also speculate the possibility of switching in topological states. The large Rashba coefficient along with topological order enables in current-induced switching in memory and logic devices, as well as magnetic field-free manipulation of spin\cite{Gupta2023}. Furthermore, this facilitates the design of high-density p–n junction arrays, potentially advancing applications in nanoelectronics and spintronics including electron-beam supercollimation\cite{Park2008,Shi2016}. 

Building upon this framework, we perform a comprehensive \textit{first-principles} calculations and symmetry analysis on the hexagonal ABC-type LiZnAs compound. The rigorous electric free energy approach was used alongside LO mode instability for the fundamental understanding of HyFE nature. Moreover, the necessity of full-potential methods accurately capturing the Rashba spin-splitting is demonstrated along with moderate strain induced TPT in LiZnAs. At the critical point, LiZnAs exhibits a Weyl semimetallic phase as well as TI phase and large Rashba coefficient after TPT. The spin texture reversal through polarization switching validate the electrical control of states. Our findings provide an arena for as-yet-unexplored LiGaGe-type materials to realize coexistence of multiple phenomena, while also establishing robust theoretical strategies for accurately predicting their complex physical properties.

\section{\label{2}COMPUTATIONAL DETAILS}

The \textit{first-principles} based density functional theory (DFT) calculations presented in this work are performed using Vienna Ab-initio Simulation Package (VASP 6.3.2)\cite{VASPKresse1996}. The contribution from the ionic potential is incorporated using projector-augmented wave (PAW) approach\cite{PAWBlochl1994}. The exchange-correlation effects are treated using the Perdew–Burke–Ernzerhof (PBE) functional within the framework of the generalized gradient approximation (GGA)\cite{GGA}. A $\Gamma$-centered k-point grids of 12~$\times$~12~$\times$~4 was adopted to integrate the Brillouin zone (BZ). The lattice parameters and atomic positions were optimized with structural relaxation until the residual forces on every atoms become less than 0.001 eV \AA$^{-1}$ within the Hellmann–Feynman criterion. The cutoff energy and convergence criteria are set to 500 eV and 10$^{-6}$ eV, respectively. The spin-orbit coupling was considered in the calculation of electronic properties to capture the Rashba effect. A detailed examination of the spin texture was conducted using a dense $15 \times 15$ $k$-point mesh in the $k_x$–$k_y$ plane, specifically centered around a high-symmetry point using PYPROCAR\cite{PYPROCAR2020}. The calculated electronic bands were projected onto a set of maximally localized Wannier functions (MLWFs) using the \textsc{Wannier90} package~\cite{WANNIER90Mostofi2014} to construct the tight-binding (TB) Hamiltonian. Subsequently, topological features such as surface states, Wilson loops, $\mathbb{Z}_2$ invariants, Weyl nodes and Fermi arcs were investigated via the iterative \textit{Green's function} method as implemented in the \textsc{WannierTools} package~\cite{WANNIERTOOLS2018}.

Our recent study observed that WIEN2k code effectively captures the potential gradient of surface asymmetric systems facilitating accurate predictions of Rashba effect\cite{Saurav2025}. In this regard, along with pseudopotential-based VASP calculations, the electronic properties of LiZnAs with SOC was also studied using the augmented plane wave plus atomic orbitals method as implemented in the full-potential WIEN2k 23.2 code\cite{WIEN2kBlaha2020}. The PBE-GGA functional was used to describe exchange-correlation effects. A $14 \times 14 \times 7$ k-point grid was adopted for Brillouin zone sampling, with muffin-tin radii fixed at 2.5 bohr. The basis set cutoff was defined via $R_\mathrm{MT} \times K_\mathrm{max} = 7.0$.

To accurately investigate the HyFE properties, we have used Quantum Espresso (QE) package\cite{QE2009} owing to well-established approach for free energy calculations under OCBC, as demonstrated in recent works by co-authors of this study\cite{Qiu2021,Qiu2023}. The norm-conserving pseudopotentials based on the Troullier-Martins scheme are employed to model core electron interactions, while the local density approximation (LDA)\cite{LDA1964} is used as exchange correlation functional since it is more efficient than GGA in phonon calculations. The single particle wavefunctions are expanded using plane waves with cutoff energy of 90 Ry and $8 \times 8 \times 4$ kpoints are considered to sample the BZ.

\section{\label{3}RESULTS AND DISCUSSION}

\subsection{Structural and Hyperferroelectric Properties of LiZnAs Compound}

\begin{figure}[H] 
	\centering
	\includegraphics[width=0.95\linewidth]{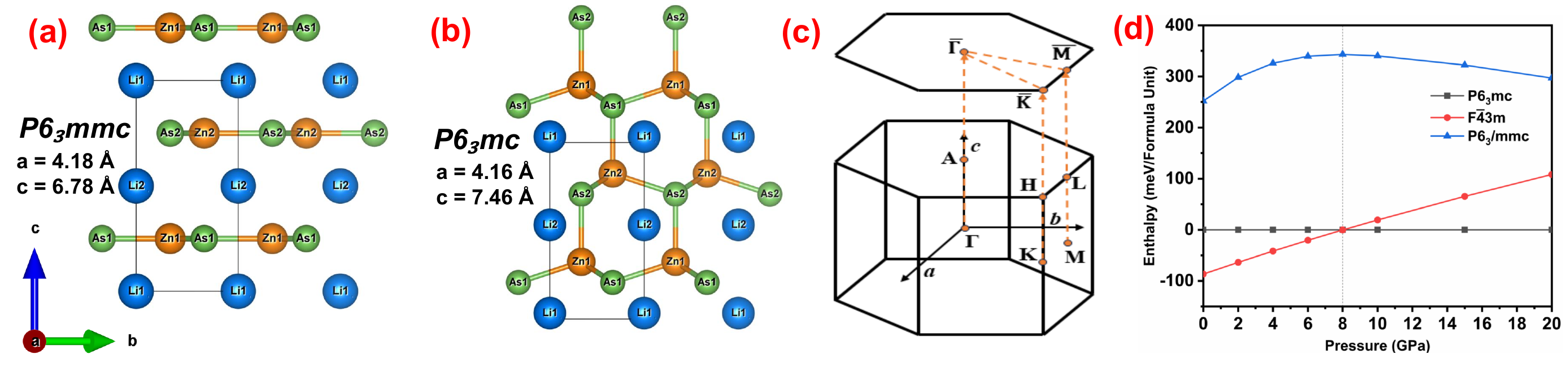}
	\caption{Structural geometry of LiZnAs compound in the centrosymmetric (paraelectric) \textit{P6$_3$/mmc} phase (a) and in the noncentrosymmetric (ferroelectric) \textit{P6$_3$mc} phase (b), respectively. The solid black line elucidates the unit cell of respective phases. The corresponding bulk and (001) surface Brillouin zones (c) and the pressure-enthalpy profile of competing \textit{P6$_3$/mmc}, \textit{P6$_3$mc} and cubic $F\overline{4}3m$ phases.}
	\label{Fig:structural+BZ}
\end{figure}

\begin{figure}[H] 
	\centering
	\includegraphics[width=0.99\linewidth]{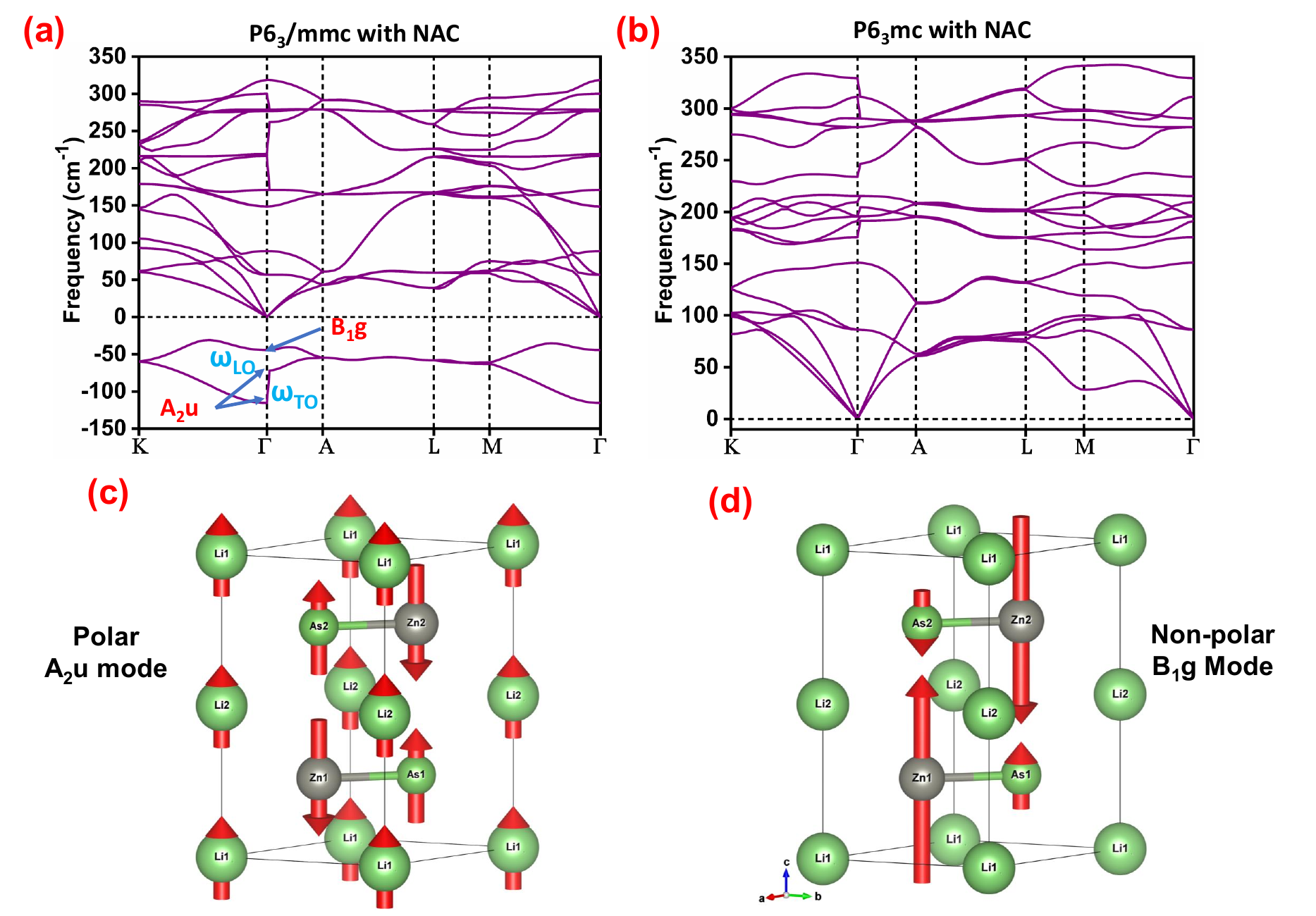}
	\caption{The phonon dispersion curves of LiZnAs compound with (a) non analytic correction (NAC) term in \textit{P6$_3$/mmc} phase and with NAC in \textit{P6$_3$mc} phase (b), respectively. The calculated phonon eigenvectors for soft longitudinal-optic $A_{2u}(\mathrm{LO})$ (c) and $B_{1g}(\mathrm{LO})$ (d) modes, respectively. The arrows indicate corresponding atomic vibrations and amplitude.}
	\label{Fig:phonons+eigenvectors}
\end{figure}

The LiZnAs compound belongs to family of Zintl phases and crystallizes in the hexagonal \textit{P6$_3$mc} (No. 186) space group exhibiting LiGaGe-type structure. This layered structure has been identified as ABC-type semiconductor and candidate HyFE in the seminal work by Bennett \textit{et al}\cite{Bennett2012}. However, the study was limited to evaluating LO mode instability as the criterion for HyFE which renders the need of rigorous electric free energy approach for further validation and fundamental understanding of HyFE properties. The non-centrosymmetric polar \textit{P6$_3$mc} phase is distorted version of high-symmetric non-polar \textit{P6$_3$mmc} (No. 194) phase. The Li$^{+}$ ions occupy the stuffing positions within the wurtzite-type [ZnAs]$^{-}$ lattice, resulting in net polarization due to potential imbalance along the crystallographic \textit{z}-axis. The structural geometry of non-polar and polar phases of LiZnAs compound is elucidated in Fig.~\ref{Fig:structural+BZ}(a) and ~\ref{Fig:structural+BZ}(b), respectively. In \textit{P6$_3$mc} phase, one can clearly observe the inversion symmetry breaking buckling of ZnAs layers. The corresponding bulk and (001) surface BZ is represented in Fig.~\ref{Fig:structural+BZ}(c). The optimized lattice constants for the polar phase using VASP package within the GGA are found to be \( a = 4.18 \, \text{\AA} \) and \( c = 6.78 \, \text{\AA} \). These values are in close agreement with those reported in prior studies by the LDA (\( a = 4.10 \, \text{\AA} \) and \( c = 6.67 \, \text{\AA} \))\cite{Bennett2012}. The minor variation arises from the inherent difference between the GGA and LDA exchange-correlation treatments. The calculated formation energy of -2.97 eV in \textit{P6$_3$mc} phase indicates the chemical stability and restrict the decomposition of LiZnAs compound in its constituent elements. A detailed examination of phonon modes offers a comprehensive perspective on lattice dynamics and their inherent dynamical stability where the instability is derived from imaginary frequencies of the soft phonon modes\cite{Gupta2013}. In this regard, the absence of imaginary frequencies over the entire BZ as depicted in Fig.~\ref{Fig:phonons+eigenvectors}(b) for \textit{P6$_3$mc} phase ensure the dynamical stability. The successful synthesis of isostructural LiGaGe-type compounds such as LiZnSb\cite{LiZnSbSynthesis2020} and LiZnBi\cite{Cao2017} suggests a viable route for experimental realization of LiZnAs in \textit{P6$_3$mc} phase. In the case of LiZnSb, epitaxial stabilization of the polar hexagonal (\textit{P6$_3$mc}) phase was achieved despite the cubic (\textit{F$\bar{4}$3m}) phase being slightly lower in formation energy, highlighting the role of kinetics to stabilize hexagonal polymorph over the cubic competing phase\cite{LiZnSbSynthesis2020}. To gain further insight, we performed total energy calculations for the experimentally observed cubic \textit{F$\bar{4}$3m} phase of LiZnAs, which is found to be energetically favorable only by 86 meV per formula unit relative to the polar hexagonal \textit{P6$_3$mc} phase. Since high-pressure–induced structural phase transition is a promising approach to overcome the energy gap, we performed volume relaxations of the known phases under pressures up to 20 GPa \cite{Zeeshan_2019, RabahKhenata2007}. As shown in the enthalpy–pressure profile Fig. \ref{Fig:structural+BZ}(d), the \textit{P6$_3$mc} phase stabilize above 8 GPa, a pressure that is experimentally accessible with current setups. The phonon dispersion curves of the \textit{P6$_3$mc} phase at 8 GPa pressure as illustrated in Fig. S1 of the Supplemental Material \cite{supplemental_material} exhibits only real frequencies, confirming the dynamical stability of this structure under elevated pressure. According to kinetics of the synthesis, this high-pressure phase can be quenched back to the ambient conditions.

To identify the rare HyFE compounds, we employed three distinct strategies aimed at validating a robust predictive approach. These methods are as follows; (i) The \textit{linear-response density functional perturbation theory} (DFPT)\cite{DFPTRevModPhys.73.515} was used to calculate the phonon eigenvalues and eigenvectors by solving secular equation, $\det\left| \frac{1}{\sqrt{M_i M_j}} C_{i\alpha, j\beta}(\mathbf{q}) - \omega^2(\mathbf{q}) \right| = 0$ where \( C(\mathbf{q}) \) is the force-constant matrix, \( M_i \) and \( M_j \) are atomic masses, \( \omega \) is the phonon eigenfrequency and \( \mathbf{q} \) is the phonon wavevector. The matrix \( C(\mathbf{q}) \) is obtained as a sum of the analytic (\( C^{\mathrm{a}} \)) and nonanalytic (\( C^{\mathrm{na}} \)) parts, respectively. The former is applicable when no macroscopic field is present, i.e., under short-circuit boundary conditions (SCBC) and can be determined from DFPT calculations. The latter originates from long-range Coulomb interactions between lattice vibrations and the macroscopic electric field, which become significant under OCBC\cite{Qiu2023}. The term \( C^{\mathrm{na}} \) is calculated from the Born effective charges (\( \mathbf{Z} \)), the high-frequency dielectric tensor (\( \boldsymbol{\varepsilon}^{\infty} \)) due to the electronic response and the unit cell volume (\( \Omega \)) using the relation \( C^{\mathrm{na}}_{i\alpha,j\beta}(\mathbf{q}) = \dfrac{4\pi e^2}{\Omega} \dfrac{(\mathbf{q} \cdot \mathbf{Z}_i)_\alpha (\mathbf{q} \cdot \mathbf{Z}_j)_\beta}{\mathbf{q} \cdot \boldsymbol{\varepsilon}^{\infty} \cdot \mathbf{q}} \)\cite{Adhikari2019}. It is worthy to note that the soft TO mode indicates the ferroelectric nature under SCBC while the non-analytic term induced soft LO mode with LO-TO splitting is prerequisite (although not sufficient) for the HyFE nature. (ii) In a ferroelectric material subjected to an OCBC, the electric displacement field satisfies \( D = 0 \). Under this constraint, a macroscopic electric field \( E \) naturally arises within the crystal. This internal field interacts with the intrinsic polarization of the system and leads to a redistribution of the electronic charge density by modifying the electron wavefunctions. For the existence of HyFE nature, by definition, material must persist spontaneous polarization at free energy minimum under OCBC. To validate this, we have used the rigorous electric free-energy framework, in which after the identification of dynamically unstable LO mode in high-symmetry non-polar paraelectric (PE) phase, the atomic configuration is systematically displaced along its normalized eigendisplacement vector $|\mathbf{u}_i\rangle$. This structural evolution is parametrized by $\lambda$ as $\mathbf{r}_i(\lambda) = \mathbf{r}^{\text{PE}}_i + \lambda c\, \mathbf{u}_i$, where $\mathbf{r}^{\text{PE}}_i$ represents atomic positions in the centrosymmetric PE phase and $c$ is the lattice constant along the polar axis. The path traced by varying $\lambda$ provides intermediate structures between PE $(\#194)$ and FE $(\#186)$ phases owing to structural phase transition. By imposing the condition of zero electric displacement (\( D = 0 \)) characteristic of OCBC, the electric free energy evaluated along the structural distortion path defined by the soft LO phonon mode can be analytically expressed as \cite{Adhikari2019, Qiu2021}; 

\begin{equation}
	F(\lambda) = U(\lambda) + \Omega(\lambda) \cdot \frac{1 + \frac{1}{2} \chi_\infty(\lambda)}{\varepsilon_0 [1 + \chi_\infty(\lambda)]^2} \cdot P^2(\lambda),
	\label{eq:FreeEnergy_OCBC}
\end{equation}

where $U(\lambda)$ is internal energy, $P(\lambda)$ represents electric polarization, $\chi_\infty(\lambda) = \varepsilon_\infty^{33}(\lambda) - 1$ is component of high-frequency dielectric permittivity and $\Omega(\lambda)$ indicates unit-cell volume, respectively. The second term in equation (1) quantifies the depolarization energy $U_{\text{dp}}(\lambda)$, which opposes the polarization under OCBC. Hence, if the free energy $F(\lambda)$, attains its global minimum at a finite distortion amplitude $\lambda \neq 0$ indicating a spontaneously polarized state stable even in the absence of screening indicating HyFE nature. (iii) The electric polarization \( P \) appearing in equation (1) is evaluated using the modern theory of polarization based on the geometric Berry-phase formalism\cite{King-Smiththeoryofpolarisation1993}. Deeper insight into the microscopic origin of electric polarization is provided in Ref.~\cite{Yao2009polarisationindepth}.

\begin{figure}[H] 
	\centering
	\includegraphics[width=0.95\linewidth]{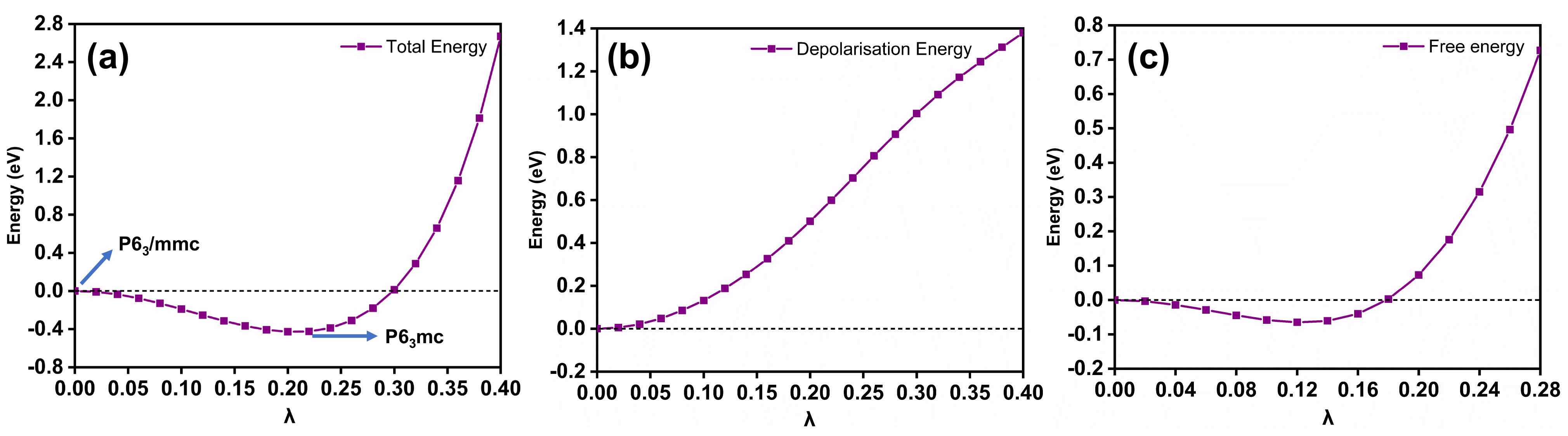}
	\caption{The calculated total internal energy $U(\lambda)$ (a), depolarization energy $U_\mathrm{dp}(\lambda)$ (b), and total free energy $F(\lambda)$ (c) as a function of distortion $\lambda$ corresponding to all the intermediate structures along the distortion pathway from the high-symmetry \textit{P6$_3$/mmc} phase to the low-symmetry \textit{P6$_3$mc} phase.}
	\label{Fig:free energy approach}
\end{figure}

The optimized lattice constants of PE LiZnAs phase are  \( a = 4.16 \, \text{\AA} \) and \( c = 7.46 \, \text{\AA} \). It is worthy to note that modern theory of polarization via Berry phase method valid only when the material is insulating within the adiabatic evolution of structure from PE to FE phase. The insulating nature of both polar and non-polar phases are determined by the calculation of electronic band structure over the BZ. The absence of electronic states over the Fermi level in the entire BZ as depicted in Fig. S2(a and b) of the Supplemental Material\cite{supplemental_material} confirms the insulating nature of both phases. The calculated electric polarization of LiZnAs compound under SCBC was found to be 0.76 $C/m^{2}$, which is in well agreement with prior study (0.75 $C/m^{2}$)\cite{Bennett2012}. The optical phonons at the $\Gamma$-point in non-polar \textit{P6$_3$mmc} phase from group theoretical analysis yields the irreducible representations (IRs) as;  

\begin{equation}
	\Gamma_{\text{Opt}}^{\text{P}6_3/\text{mmc}} = 2E_{2g} \oplus 2E_{1u} \oplus 2A_{2u} \oplus E_{2u} \oplus 2B_{1g} \oplus B_{2u}
\end{equation}
where all \textit{E} modes are doubly degenerate while \textit{A} and \textit{B} are singly degenerate. In contrast, zone-center IRs of polar \textit{P6$_3$mc} phase are classified as; 

\begin{equation}
	\Gamma_{\text{Opt}}^{\text{P}6_3\text{mc}} = 3E_{2} \oplus 2E_{1} \oplus 2A_{1} \oplus 3B_{1}.
\end{equation}
The phonon dispersion curves using \textit{linear-response} without incorporating non-analytic term correction (\( C^{\mathrm{na}} \)) exhibits two imaginary (dynamically unstable) modes; the polar $A_{2u}$ and non-polar $B_{1g}$ modes at frequencies 115.63\textit{i} $cm^{-1}$ and 44.46\textit{i} $cm^{-1}$, respectively. The eigenvectors corresponding to unstable $A_{2u}$ and $B_{1g}$ modes are depicted in Fig.~\ref{Fig:phonons+eigenvectors}(c and d), respectively. In the case of $A_{2u}(TO)$ mode, we observed that Zn and As atoms vibrate in opposite directions along the $c$-axis with notable amplitudes, indicating a polar character with minimal Li contribution. In contrast, $B_{1g}$ mode exhibits anti-parallel $Zn-Zn$ and $As-As$ atomic vibrations within the primitive cell. These anti-symmetric displacements cancel the macroscopic dipole moment, resulting in a non-polar vibration character. The Li atoms remain effectively immobile, indicating their negligible participation in this mode. To investigate the HyFE nature under OCBC, the non-analytic long-range Coulomb term (\( C^{\mathrm{na}} \)) was included in the dynamical matrix which governs the LO-TO splitting\cite{Li2016}. In general, conventional FEs such as PbTiO$_3$ and BaTiO$_3$ have large Born effective charges combined with small high-frequency dielectric constants ($\varepsilon_\infty$) result in significant LO-TO splitting. This ensures that all LO modes have real frequencies indicating the unscreened depolarization field and no spontaneous polarization. In sharp contrast, HyFE materials have small LO-TO splitting with unstable LO and TO modes which reinforce a robust polarization state that persists despite presence of depolarization field. As illustrated in Fig.~\ref{Fig:phonons+eigenvectors}(a), polar $A_{2u}(LO)$ mode remains soft after LO-TO splitting with frequency $71.86\textit{i}$ $cm^{-1}$, satisfying the prerequisite for HyFE nature while the non-polar $B_{1g}$ mode retains its frequency owing to the absence of interactions with macroscopic electric field. Moreover, it is important to emphasize that the low-energy polar \textit{P6$_3$mc} phase exhibits real phonon frequencies over the entire BZ, confirming its dynamical stability as shown in Fig.~\ref{Fig:phonons+eigenvectors}(b). To substantiate the presence of HyFE in LiZnAs, we evaluate the electric free energy ($F$) along the configuration path defined by the eigendisplacement vector corresponding to the unstable LO phonon mode.
The single-point energy calculations are carried out on a series of intermediate structures to construct the potential energy surface as a function of the distortion $(\lambda)$ according to unstable LO mode. The calculated internal energy $U(\lambda)$, depolarization energy $U_\mathrm{dp}(\lambda)$ and total free energy $F(\lambda)$ corresponding to the intermediate structures along the distortion pathway from the high-symmetry \textit{P6$_3$/mmc} phase to the low-symmetry \textit{P6$_3$mc} phase are presented in Fig.~\ref{Fig:free energy approach}(a-c). In particular, Fig.~\ref{Fig:free energy approach}(a) depicts the structural transition towards lower energy polar phase as $\lambda$  deviates from zero. According to equation (1), inclusion of depolarization energy as shown in Fig.~\ref{Fig:free energy approach}(b) leads to the total free energy under OCBC. The electric free energy $F(\lambda)$ decreases with increasing deviation of $\lambda$ from zero (see Fig.~\ref{Fig:free energy approach}(c)), reaches a minimum at $\lambda = 0.12$ and subsequently increases.  The structure corresponding to $\lambda = 0.12$ is thus identified as the LO-mode-driven HyFE phase owing to existence of non-zero polarization at minimum of $F(\lambda)$ under OCBC. 

One of the key limitations in previously explored HyFE materials is the remarkably shallow depth of the electric free energy well under OCBC, which critically undermines their thermal robustness. As a result, thermal fluctuations at even modest temperatures can readily suppress the polar phase, making HyFE behavior observable only at cryogenic temperatures. It is therefore of paramount importance to identify systems where the HyFE free energy landscape exhibits a significantly deeper minimum. In this regard, the free energy well depth of LiZnAs compound is found to be $\Delta F_{\mathrm{HyFE}} = -66$ meV which is significantly higher than previously reported LiNbO$_3$ with merely $-9$~meV well depth\cite{Qiu2021}. Hence, \textit{LiZnAs emerges as a rare example of an intrinsic HyFE material with a substantial free energy well depth, underscoring its fundamental significance in the evolving field of Hyperferroelectricity}. We have also calculated the polarization of HyFE phase under OCBC, using Berry-phase approach\cite{King-Smiththeoryofpolarisation1993}. The HyFE ground state with an optimal displacement amplitude of $\lambda = 0.12$ yields a polarization of $P_\mathrm{HyFE} = 0.282~\mathrm{C/m^2}$, which is \textit{giant} and nearly an order of magnitude larger than the $\sim 0.023~\mathrm{C/m^2}$ values of $P_\mathrm{HyFE}$ reported in LiBeSb\cite{Bennett2012} and LiNbO$_3$\cite{Qiu2021} compounds. Remarkably, the obtained $P_\mathrm{HyFE}$ under OCBC surpasses the spontaneous polarization of prototypical ferroelectric BaTiO$_3$ ($\sim 0.21~\mathrm{C/m^2}$) under SCBC, highlighting the exceptional nature of HyFE despite presence of depolarization field in LiZnAs compound. One can speculate that the large HyFE polarization facilitates the control over transport properties in superconductors and topological materials via formation of interfaces or heterostructures\cite{Qiu2023} along with complimentary ability to polarize in ultrathin layers which may allow ultrafast switching behaviour\cite{Garrity2014}.

The stabilization of polarization under OCBC in LiZnAs calls for an examination of the microscopic origin of its intrinsic HyFE nature. It is imperative to emphasize that the HyFE polarization can be attributed to the mode effective charge (MEC) associated with soft LO phonon and high-frequency electronic dielectric constant. The first quantity controls the strength of coupling between lattice vibrations and macroscopic polarization, while the second provides dielectric screening that reduces the energy cost of the depolarization field under OCBC. In general, for a phonon branch $n$ at wave vector $\mathbf{q}$ with normalized eigendisplacements $u^{(n\mathbf{q})}_{i\beta}$, the \textit{mode-specific effective charge} is defined as;\cite{Shao2021} 
\begin{equation}
	\tilde{Z}^{(n\mathbf{q})}_{\alpha} = \sum_{i\beta} 
	Z^{*}_{i,\alpha\beta}\, u^{(n\mathbf{q})}_{i\beta},
\end{equation}
where $Z^{*}_{i,\alpha\beta}$ is the Born effective-charge tensor of atom $i$ while $\alpha$ and $\beta$ denote Cartesian directions, and the summation extends over all atoms $i$ and coordinates $\beta$. It is worthy to note that $\tilde{Z}^{(n\mathbf{q})}_{\alpha}$ is now a vector which specifies the polarization induced by the collective motion of the mode. In this regard, HyFE polarization is directly proportional to the MEC. For a phonon-induced structural distortion of amplitude $\lambda$ along the polar $c$-axis, one obtains
\begin{equation}
	\Delta P_{\alpha} = \frac{1}{\Omega}\, \tilde{Z}^{(n\mathbf{q})}_{\alpha}\, c\,\lambda,
\end{equation}
where $\Omega$ is the unit cell volume and $c$ is the lattice constant along the polarization axis. A larger MEC therefore leads to a larger HyFE polarization. In practice, the most relevant contribution comes from the $\tilde{Z}_3$ component owing polar $c$-axis, which dominates the stability of hyperferroelectricity in LiZnAs.The calculated $\tilde{Z}_3$ component for the soft $A_{2u}(\mathrm{LO})$ mode in intrinsic LiZnAs is found to be $4.8364$. This value is notably larger than the MEC reported for LiNbO$_3$ ($\tilde{Z}_3 \approx 0.39$) \cite{Qiu2021} and LiZnSb under $-2\%$ inplane strain ($\tilde{Z}_3 \approx 2.59$) \cite{Qiu2023}, validating stronger HyFE nature in LiZnAs. Our findings also validate that larger MECs and dielectric constant contribute to giant hyperferroelectricity. Notably, small band gap materials facilitates large dielectric constants. 

\subsection{Electronic Properties of LiZnAs Compound}
Exploiting electronic degrees of freedom is central to advancing the spintronics and low-power electronics\cite{Bahramy2012}. In particular, the interplay of Rashba spin-splitting and non-trivial band topology within a single-material platform offers a promising avenue. In the presence of inversion symmetry breaking, a spin-degenerate parabolic band splits into two spin-polarized branches, governed by the dispersion relation \( E_{\pm}(\mathbf{k}) = \frac{\hbar^2 k^2}{2m^*} \pm \alpha_R |\mathbf{k}| \), where \( \alpha_R \) is the Rashba coefficient. This splitting results in a momentum offset \( (\Delta k) \) and Rashba splitting energy \( (\Delta E) \), which are related to Rashba coefficient as $\alpha_R = 2 \Delta E / \Delta k$. Consequently, two Rashba-split bands possess opposite spin orientations, which directly affect the system’s magnetic and optical response under photoexcitation. This spin–momentum locking enhances spin-to-charge conversion efficiency, making it attractive for spintronic applications\cite{Manchon2015, Lesne2016}. To investigate the electronic properties, the crystal structure corresponding to the minimum of the potential energy surface (characterized by broken inversion symmetry) was fully relaxed which results in the energetically favored polar $P6_3mc$ phase. 

\begin{figure}[H] 
	\centering
	\includegraphics[width=0.98\linewidth]{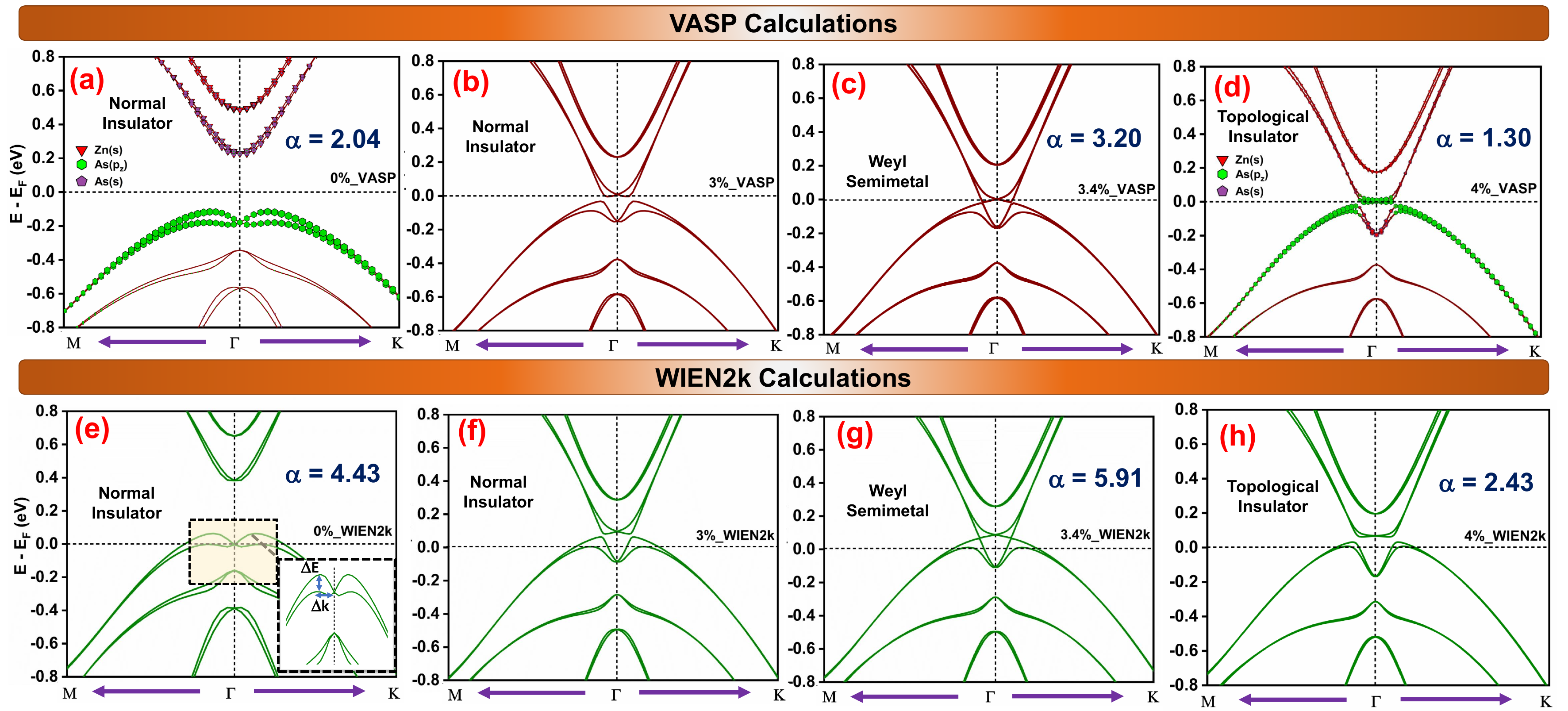}
	\caption{The calculated relativistic band structure of polar \textit{P6$_3$mc} phase using VASP (top panel) and WIEN2k (bottom panel). The evolution of electronic bands around the Fermi level under biaxial strain along M--$\Gamma$--K high symmetry path from VASP (a-d) and from WIEN2k (e-h), respectively. The similar nature of bands dispersion with significantly distinct Rashba coefficient was evidenced directly before and after the topological phase transition. The symbols in (a and d) mimics the corresponding orbital contribution}.
	\label{Fig:electronic band structures}
\end{figure}

The calculated relativistic band structure for this phase as illustrated in Fig.~S2(b) of the Supplemental Material\cite{supplemental_material} shows prominent RSS at generic momenta, except along the $\Gamma$--$A$ direction and at the time-reversal invariant momenta (TRIMs) points, where symmetry constraints suppress the splitting. The symmetry operations of the $P6_3mc$ space group includes
$\{E\,|\,000\}$, $\{C_2\,|\,00\tfrac{1}{2}\}$, $\{C_6^+\,|\,00\tfrac{1}{2}\}$, $\{C_6^-\,|\,00\tfrac{1}{2}\}$, 
$\{C_3^+\,|\,000\}$, $\{C_3^-\,|\,000\}$, $\{\sigma_{v1}\,|\,000\}$, $\{\sigma_{v2}\,|\,000\}$, 
$\{\sigma_{v3}\,|\,000\}$, $\{\sigma_{d1}\,|\,00\tfrac{1}{2}\}$, $\{\sigma_{d2}\,|\,00\tfrac{1}{2}\}$, and 
$\{\sigma_{d3}\,|\,00\tfrac{1}{2}\}$, respectively\cite{Yu2022Encyclopedia}.
Specifically, along the $\Gamma$--$A$ path, the wave vector is invariant under the little group isomorphic to $C_{6v}$, whose non-commuting $C_6$ and $\sigma_v$ operations enforce double degeneracy, prohibiting RSS. In contrast, the centrosymmetric $P6_3/mmc$ phase (see Fig.~S2(a) of the Supplemental Material\cite{supplemental_material}) retains spin degeneracy across all bands due to preserved inversion symmetry. In principle, the RSS is maximized in momentum planes that are orthogonal to the direction of the polar axis. Since the polar axis in LiZnAs aligns with the crystallographic $c$ direction, the RSS was found to be pronounced in the $M$--$\Gamma$--$K$ plane. In this regard, the electronic band structure with SOC using VASP and WIEN2k along the  $M$--$\Gamma$--$K$ direction is illustrated in Fig.~\ref{Fig:electronic band structures}(a and e), respectively. It is widely recognized that orbital band inversion around the Fermi level is the hallmark of TIs. As evident from Fig.~\ref{Fig:electronic band structures}(a), the \( s \) orbital of Zn and As atoms together with \( p_z \) orbital of As atom contribute significantly in the conduction band minima and valence band maxima, respectively. The normal band ordering indicates the trivial insulating nature of LiZnAs compound with pronounced RSS under ambient conditions. Notably, the electronic structure of crystalline materials is strongly tied to the hybridization between atomic orbitals, which can be systematically tuned by external parameters such as pressure, doping, or alloying. Among others, in-plane biaxial strain (BAS) has emerged as a particularly effective and clean tuning knob for driving TPTs, offering an advantage over chemical doping by avoiding disorder and compositional inhomogeneities\cite{Barman2018}. In this study, we induce TPT via moderate BAS applied along the crystallographic [110] direction, achieved through controlled lattice deformation. This strain engineering strategy mimics the lattice mismatch effect commonly realized in epitaxial growth on mismatched substrates, thereby providing an experimentally feasible route to strain-driven TPTs.

The calculated Rashba energy splitting \( (\Delta E) \) and momentum offset \( (\Delta k) \) using VASP and WIEN2k is significantly different which  leads to different $\alpha_R$, as summarized in Table~\ref{Tab:Rashba parameters} despite exhibiting similar nature of band dispersion. It is worth highlighting that the critical role of full-potential code as compared to pseudopotential code for accurate prediction of intrinsic RSS was discussed in our recent work with physical mechanism that causes the difference in calculated $\alpha_R$\cite{Saurav2025}. \textit{This prompts a fundamental question that does the same mechanism hold when the TPT is triggered by applied strain in surface asymmetric compounds? Does the critical point of TPTs shifts?}. Let's understand this from subsequent discussion of bands around Fermi level. The evolution of band structure with BAS strain is elucidated in Fig.~\ref{Fig:electronic band structures} using VASP and WIEN2k, respectively with corresponding Rashba parameters in Table~\ref{Tab:Rashba parameters}. Therefore, full-potential based WIEN2k calculations are more accurate for the investigation of $\alpha_R$\cite{Saurav2025}. It is well established that the TPT between a non-centrosymmetric TIs and a trivial insulator inevitably involves an intermediate Weyl semimetal phase\cite{Liu2014, Murakami2008}. Consequently, the electronic bands with increase in BAS decreases the band gap which leads to the quantum critical point around the moderate 3.4\% strain in the LiZnAs HyFE Rashba semiconductor. At this strain, the system enters a Weyl semimetallic phase, characterized by Weyl nodes along both $M$--$\Gamma$ and $\Gamma$--$K$ directions, as evidenced using both VASP (Fig.~\ref{Fig:electronic band structures}(c)) and WIEN2k (Fig.~\ref{Fig:electronic band structures}(g)) calculations. Beyond the critical point, BAS reopens the band gap with the inverted band ordering between \( s \) orbital of Zn and As atoms together with \( p_z \) orbital of As atom as shown in  Fig.~\ref{Fig:electronic band structures}(d) for 4\% BAS confirming the TPT in LiZnAs from trivial insulator to topological insulator. Moreover, our calculations upto the 6\% BAS confirms the persistant orbital band inversion post-topological phase transition. The polar phase remains energetically favorable across this moderate strain-induced TPT, highlighting the inherent stability and resilience over centrosymmetric non-polar phase. In the case of RSS, a giant Rashba coefficient of 5.91 eV $\text{\AA}$ is observed for the Weyl semimetal phase while the topological insulator phase at 4\% BAS exhibits $\alpha_R$ = 2.43 eV $\text{\AA}$ as the band gap reopens across the transition. This further increases to 3.69 eV $\text{\AA}$ under $6\%$ biaxial strain (band dispersion not shown here). Our findings emphasizes the tunability of spin-splitting in polar Rashba semiconductors and its intimate connection with the underlying multiple topological phases. The calculated values of $\alpha_R$ are significantly larger or comparable to well known BiTeI ($\alpha_R$ = 3.8 eV \text{\AA})\cite{Ishizaka2011}, BiTeCl ($\alpha_R$ = 1.20 eV \text{\AA})\cite{XiangBiTeCl2015}, Bi on Ag surface alloy ($\alpha_R$ = 3.05 eV \text{\AA})\cite{Ast2007}, KSnBi ($\alpha_R$ = 1.17 eV \text{\AA})\cite{Mondal2021}, Sb$_{2}$TeSe$_{2}$ ($\alpha_R$ = 3.88 eV \text{\AA})\cite{MeraAcosta2020}, KMgSb ($\alpha_R$ = 0.83 eV \text{\AA}), LiZnSb ($\alpha_R$ = 1.82 eV \text{\AA}) and LiCaBi ($\alpha_R$ = 1.82 eV \text{\AA})\cite{Narayan2015} among others. Despite the well-known band gap underestimation associated with the GGA-PBE functional, its reliability in capturing Rashba-type spin-splitting is well established. This was evidenced by the strong agreement between previously reported full-potential WIEN2k-based predictions and experimental observations for systems such as bulk BiTeI\cite{Ishizaka2011} and the Pb-induced Ge(111) surface\cite{Yaji2010}. This consistency underscores the suitability of this functional for the present investigations. 

It is worthy and interesting to note that, \textit{while strain does not substantially alter the qualitative band evolution between full-potential WIEN2k and pseudopotential based VASP (PAW) codes, the variation in $\alpha_R$ was persistent owing to the difference lying in physical mechanism to treat the potential gradient of a crystalline material as discussed in our recent study}\cite{Saurav2025}. This insight decouples the mechanism of TPT from the intrinsic code-dependent differences in Rashba splitting, providing a deeper understanding of strain-induced topological phenomena in polar semiconductors.

\subsection{Topological Properties of LiZnAs Compound}

Although orbital band inversion alone does not constitute a rigorous criterion for defining a topological phase, it remains a valuable heuristic for screening candidate materials that may host non-trivial topological character. To substantiate the bulk-boundary correspondence of topological material, comprehensive investigation of $\mathbb{Z}_2$ topological invariants, the presence of mass-less fermionic surface states, and the analysis of the flow of Wannier charge centers (WCC) across time-reversal invariant planes are essential. The tight-binding Hamiltonian was constructed using maximally localized Wannier functions (MLWFs) obtained via the \texttt{Wannier90} package. The reliability of this Hamiltonian was validated by the direct comparison between Wannier-interpolated and DFT band dispersion indicating consistency and well-localized Wannier functions. This Hamiltonian was subsequently used to explore the topological characteristics of the system by iterative Green's function formalism as implemented in \texttt{WannierTools}. Moreover, to confirm the Weyl semimetal phase, the Berry curvature is calculated using \texttt{Wannier90}. The lack of inversion symmetry in the LiZnAs compound renders the parity-based approach of Fu and Kane\cite{Fu2007}, typically used to determine $\mathbb{Z}_2$ topological invariant at TRIM points, inapplicable. Consequently, we adopt the WCC evolution method to evaluate the topological character of the band structure. The WCC represents the mean position of charge associated with a given Wannier function and its variation as a function of crystal momentum is governed by the phase factor $\theta$ stemming from the eigenvalues of projected position operato\cite{Juneja2018}. In three-dimensional systems, the $\mathbb{Z}_2$ indices are denoted as $(\nu_0; \nu_1\nu_2\nu_3)$, where $\nu_0$ is the strong topological index and the triplet $(\nu_1, \nu_2, \nu_3)$ correspond to the weak indices. These invariants are determined by analyzing the winding behavior of WCCs over time-reversal invariant planes in the BZ. The equation for the same is given as;

\begin{equation}
	\nu_0 = \left[ \mathbb{Z}_2(k_i = 0) + \mathbb{Z}_2(k_i = 0.5) \right] \bmod 2
\end{equation}
\begin{equation}
	\nu_i = \mathbb{Z}_2(k_i = 0.5) \quad \text{for} \quad i = 1, 2, 3
\end{equation}.

\begin{figure}[H] 
	\centering
	\includegraphics[width=1.0\linewidth]{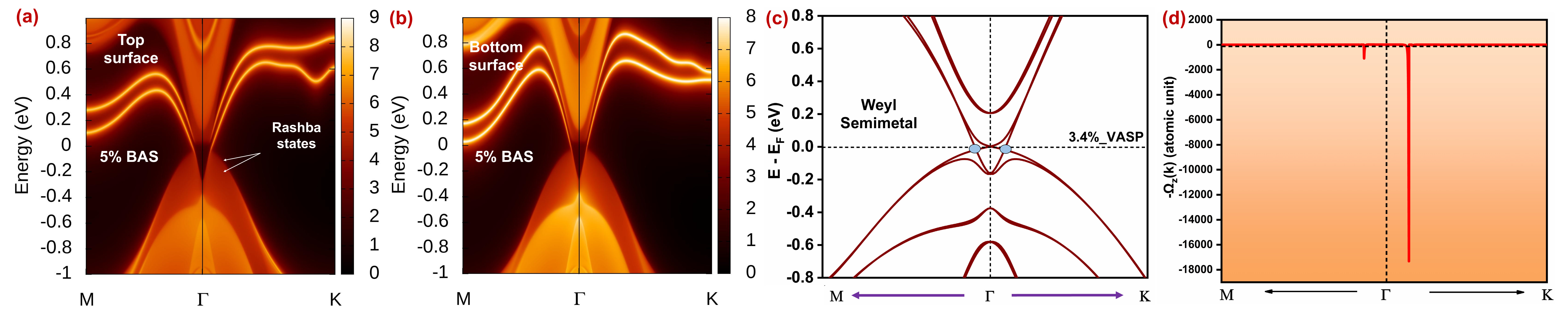}
	\caption{The calculated topological surface states at 5\% biaxial strain for top and bottom terminations of LiZnAs slab (a and b). The electronic band structure for Weyl semimetal phase at 3.4\% strain with highlighted nodes (c) and corresponding berry curvature (d), respectively.}
	\label{Fig:topological parameterts + Berry phase calculations}
\end{figure}

An odd number of crossings between the reference line and the WCC trajectories indicates a strong ($\mathbb{Z}_2 = 1$) topological phase. The evolution of WCCs was analyzed (plot not shown here) on six TRIM planes at $k_i = 0$ and $k_i = 0.5$ ($i = x, y, z$) for the system under 5\% BAS after TPT. Notably, after the 3.4\% strain, the inverted band ordering is persistent up to 6.5\% strain. However, for distinguishable observation of topological surface states we have simulated the slab for 5\% BAS.  The reference line intersects the WCC paths an odd number of times for $k_i = 0$ and an even number for $k_i = 0.5$ ($i = x, y, z$), thereby yielding the topological indices $(\nu_0; \nu_1\nu_2\nu_3) = (1;000)$ using equations (6) and (7). This confirms the emergence of a strong topological insulating phase in LiZnAs following the TPT within its non-centrosymmetric polar HyFE Rashba semiconducting regime. The calculated Dirac surface states for slab geometry of LiZnAs compound are depicted in Fig.~\ref{Fig:topological parameterts + Berry phase calculations}(a and b) corresponding to the top and bottom surfaces along the (001) termination of hexagonal BZ. It is evident that topological surface states resides within the bulk energy gap crossing the Fermi level at the zone center. Notably, the Dirac points are buried deep within the valence band, embedded in an energy valley that originates from RSS near the valence band maximum. The similar characteristics have been observed in Bi$_2$Te$_3$, KSnBi, BiTeI, CsSnBi, LiCaBi, XGeBi (X = K, Rb, Cs) and KMgBi~\cite{DiSante2016, Mondal2021, Saurav2025, Patel2024a, Bahramy2012, Gupta2023}. Furthermore, Bahramy \textit{et al.} \cite{Bahramy2012} demonstrated that in such non-centrosymmetric systems, the nature of surface states is sensitive to slab termination, leading to an asymmetric charge distribution across the top and bottom surfaces. These distinct features may offer promising avenues for exploiting spin-momentum-locked states in next-generation spintronic devices.  To substantiate the realization of the Weyl semimetal phase in LiZnAs under 3.4\% BAS, we performed a detailed analysis of the Berry curvature along the $M$--$\Gamma$ and $\Gamma$--$K$ BZ directions. The calculated band structure, shown in Fig.~\ref{Fig:topological parameterts + Berry phase calculations}(c), reveals the presence of Weyl nodes, while the corresponding Berry curvature distribution in Fig.~\ref{Fig:topological parameterts + Berry phase calculations}(d) exhibits pronounced peaks at these nodes. This behavior is a hallmark of topologically non-trivial Weyl fermions which mimics the monopole-like behaviour of Berry flux near the crossing points. Experimental verification of the non-trivial topological character of a material is typically achieved through techniques such as angle-resolved photoemission spectroscopy (ARPES), magneto-transport measurements, and analysis of quantum oscillations in electrical resistance as a function of an applied magnetic field.

\subsection{Rashba Spin texture and Switching Barrier of LiZnAs Compound}

\begin{figure}[H] 
	\centering
	\includegraphics[width=0.95\linewidth]{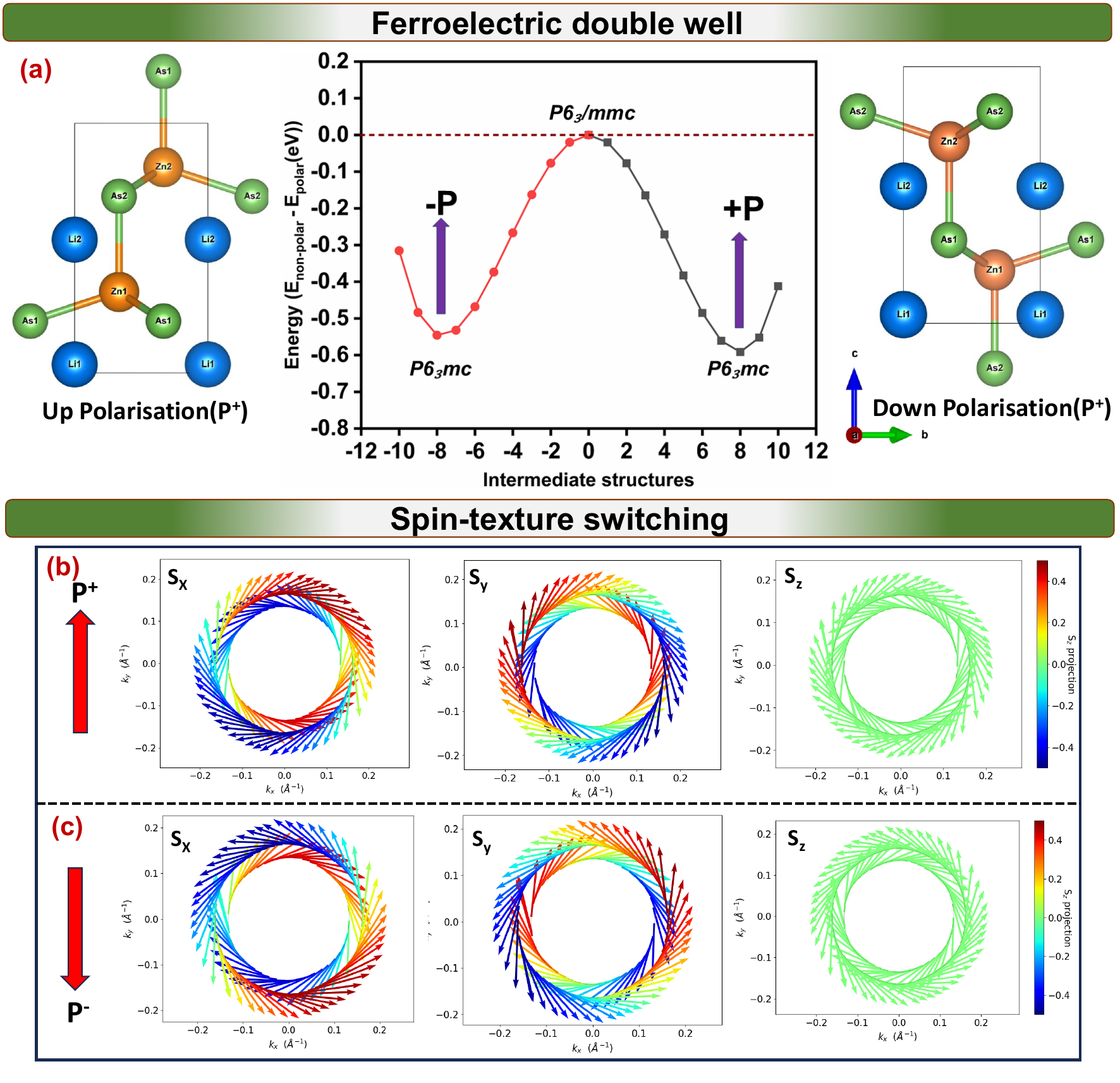}
	\caption{The characteristic ferroelectric double-well energy profile of LiZnAs compound as a function of structural phase transitions from non-polar \textit{P6$_3$/mmc} phase to polar \textit{P6$_3$mc} phase for both up and down polarization directions (a). The red and black symbols indicate the energy differences for intermediate structures. The calculated Rashba spin textures at the constant energy contour in electronic bands at -0.18 eV, highlighting the projected spin components $S_{x}$, $S_{y}$ (in-plane) and $S_{z}$ (out-of-plane) for up polarization (b) and reversed Rashba spin textures after switching in buckling of planes (down polarization) (c) in LiZnAs compound, respectively.}
	\label{Fig:double well and spin-texture switching}
\end{figure}

The dispersion of Kramers doublets near the $\Gamma$ point is dictated by the symmetry of crystal momentum $\mathbf{k}$ in the crystalline materials. For spin-1/2 fermions under the $P6_3mc$ phase, rotational symmetry operations such as $C_n$ take the form $e^{-i\sigma_z \pi/n}$, while mirror operations are described by $M_v = i\sigma_x$ and $M_d = i\sigma_y$ respectively, with $\sigma_{x,y,z}$ denoting the Pauli matrices. When combined with time-reversal symmetry $T = i\sigma_y K$, these constraints lead to the $\mathbf{k} \cdot \mathbf{p}$ Hamiltonian in the vicinity of $\Gamma$ point as:
\begin{equation}
	H_\Gamma = \frac{k_x^2 + k_y^2}{2m^*_{xy}} + \frac{k_z^2}{2m^*_z} + \alpha_R(k_x \sigma_y - k_y \sigma_x),
\end{equation}
where $\alpha_R$ quantifies the strength of Rashba coupling while $m^*_{xy}$ and  $m^*_z$ are the effective masses in respective directions. For a given electronic eigenstate, the spin component ${\sigma}$ is orthogonal to the crystal momentum $\mathbf{k}$, reflecting a pronounced spin--momentum locking characteristic. The linear term results in an in-plane spin-momentum locking, while higher-order spin warping is symmetry-forbidden by $C_2$ and $C_6$ rotations\cite{DiSante2016}. To further corroborate this, we investigated the spin textures of LiZnAs using expectation values of $\langle S_\alpha(\mathbf{k}) \rangle = \langle \psi(\mathbf{k}) | \sigma_\alpha | \psi(\mathbf{k}) \rangle$, where $\psi(\mathbf{k})$ are the spinor wavefunctions. The in-plane $\mathbf{k}$-grid of $15 \times 15$ was used within the \textsc{PYPROCAR} tool for calculations at constant energy countours. The spin texture of LiZnAs at $-0.18$~eV reveals a strictly in-plane orientation, where $S_x$ and $S_y$ components dominate while $S_z$ is effectively zero as shown in Fig.~\ref{Fig:double well and spin-texture switching}(b), which is obvious from equation (8). Notably, the Rashba-split bands show opposite orientations; counterclockwise rotation for the inner band and clockwise for the outer band, maintaining the time-reversal symmetry where $\mathbf{S}(\mathbf{p}_\parallel) = -\mathbf{S}(-\mathbf{p}_\parallel)$ and $E(\mathbf{p}_\parallel) = E(-\mathbf{p}_\parallel)$. Interestingly, reversing the bulk polarization by changing associated electric potential gradient and the spin-orbit field in LiZnAs compound results in spin texture reversal with strong in-plane spin-momentum locking as illustrated in Fig.~\ref{Fig:double well and spin-texture switching}(c). This behavior can be attributed to a sign reversal of the Rashba coefficient ($\alpha_R$). This polarization dependent spin texture switching observed in LiZnAs, demonstrates that the ferroelectric Rashba semiconductors with topological surface states offer a platform where spin degrees of freedom can be directly controlled via electric polarization\cite{DiSante2016, Narayan2015}. To further substantiate the polarization and switchability in LiZnAs, we computed the hallmark ferroelectric double-well energy profile along the structural path connecting two oppositely polarized states through the centrosymmetric $P6_3/mmc$ configuration, following the established methods for isostructural LiGaGe-type materials \cite{Bennett2012, Narayan2015}. First-principles based single point total energy calculations were then performed along this structural pathway. Consequently, Fig.~\ref{Fig:double well and spin-texture switching}(a) represents a characteristic double-well potential with minima corresponding to the polar $P6_3mc$ configurations and a saddle point at the non-polar structure as expected for polar materials. The calculated switching barrier is 0.54~eV per unit cell (two formula unit per unit cell) of LiZnAs, which is significantly lower than that of only reported isostructural HyFE LiZnSb (0.80~eV) and LiBeSb (0.58~eV) suggesting a relatively favorable energy cost for polarization reversal.

\section{\label{IV}CONCLUSIONS}
In summary, we present a predictive framework that unifies the exploration of hyperferroelectricity, multiple topological phases, Rashba spin-splitting, and spin texture switching within a single material, thereby advancing the broader paradigm of composite quantum compounds (CQCs) using \textit{first-principles} based DFT calculations. This approach capitalize on the interplay of crystalline symmetries, strength of SOC and vibrational dynamics of the investigated wurtzite  LiZnAs compound. Our main conclusions, capturing the essential insights of this work, are outlined as follows. 

(i) We have discussed HyFE properties using combined linear-response perturbation theory, modern theory of polarization and the electric free energy approach. The unstable $A_{2u}(LO)$ mode in centrosymmetric phase induce a free energy minimum at nonzero polarization with $P_\mathrm{HyFE} = 0.282~\mathrm{C/m^2}$ which is higher than HyFE LiNbO$_{3}$ ($P_\mathrm{HyFE} = 0.023~\mathrm{C/m^2}$) under OCBC and conventional BaTiO$_{3}$ ($P = 0.21~\mathrm{C/m^2}$) even under SCBC.

(ii) Intrinsically, LiZnAs compound exhibits giant free energy well depth of -66 meV indicating robust HyFE nature compared to previously reported shallow well depth of -9 meV in intrinsic LiNbO$_{3}$ where it is highly likely that even modest temperatures can readily suppress the polar phase. The physical origin of stable HyFE in LiZnAs is attributed to its large high-frequency dielectric constant, which stems from its small band gap nature. The emergence of giant HyFE polarization in this system is expected to unlock scarcely unexplored diverse opportunities for future HyFE based technologies.

(iii) Our results demonstrate that, although VASP and WIEN2k codes yield quantitatively different Rashba spin-splittings, the quantum critical point of the topological phase transition under external strain remains unchanged. This finding underscores that while a full-potential treatment is essential for accurately capturing relativistic spin-splittings, the overall evolution of band topology is robust against such differences. Furthermore, we obtained the Weyl semimetal (3.4\%) and topological insulating phase (after 3.4\%) under moderate biaxial strain indicating the freedom of tunability for practical applications. The calculated topological surface states, $\mathbb{Z}_2$ invariants and berry curvature at nodes confirms the non-trivial nature.

(iv) The reversal of spin texture driven by polarization switching offers a purely electrical route for manipulating and controlling spin degrees of freedom in these materials with giant Rashba Coefficient of 4.43 eV \text{\AA} intrinsically, 5.91 eV \text{\AA} for Weyl semimetal phase and 2.42 eV \text{\AA} at 4\% strain after TPT in LiZnAs compound.

This mechanism and more suitable approach pave the way for multifunctional non-volatile and electronics technologies thereby motivating the research in CQCs. We believe that rich and interesting results in our study facilitates theoretical and existential research to design composite quantum compounds which intrinsically exhibits hyperferroelectricity, giant Rashba effect, non-trivial band topology and electrical way for spin texture switching.

\begin{acknowledgments}
SP (DST/INSPIRE Fellowship/2022/IF220593) highly appreciates the Department of Science and Technology (DST), Government of India, for the Inspire Fellowship. Authors would like to acknowledge National Supercomputing Mission (NSM) for providing computational resource of ‘PARAM Porul’ at NIT Trichy, India. Authors are thankful to DST-FIST (SR/FST/PS-I/2022/230), Government of India, for financial assistance. Some of the calculations were carried out at the Interdisciplinary Centre for Mathematical and Computational Modelling at the University of Warsaw (ICM UW) under grant No. g102-2474.
\end{acknowledgments}

\bibliography{ref}

\newpage

\textbf{TABLES:}

\begin{table}[htbp]
	\centering
	\caption{\label{Tab:Rashba parameters}
		The calculated Rashba energy splitting ($\Delta E$ in meV), momentum offset ($\Delta K$ in \AA$^{-1}$) and Rashba coefficients ($\alpha_R$ in eV \AA) for pristine and strained LiZnAs compound using VASP and WIEN2k codes along with their topological nature, respectively.}
	\begin{tabular}{lcccc}
		\hline\hline
		\textbf{System} & \textbf{$\Delta E$ (meV)} & \textbf{$\Delta K$ (\AA$^{-1}$)} 
		& \textbf{\makecell{Rashba coefficient \\ \textbf{$\alpha_R$} (eV \AA)}} & \textbf{Nature} \\
		\hline\hline
		\multicolumn{5}{c}{\textbf{VASP Calculations}} \\
		\hline
		0\%$\_$LiZnAs   & 64 & 0.06 & 2.04 & Normal Insulator \\
		3.4\%$\_$LiZnAs & 67 & 0.04 & 3.20 & Weyl semimetal \\
		4\%$\_$LiZnAs   & 34 & 0.05 & 1.30 & Topological Insulator \\
		\hline
		\multicolumn{5}{c}{\textbf{WIEN2k Calculations}} \\
		\hline
		0\%$\_$LiZnAs   & 66 & 0.03 & 4.43 & Normal Insulator \\
		3.4\%$\_$LiZnAs & 70 & 0.02 & 5.91 & Weyl semimetal \\
		4\%$\_$LiZnAs   & 35 & 0.28 & 2.42 & Topological Insulator \\
		\hline\hline
	\end{tabular}
\end{table}

\newpage
\textbf{FIGURES:}

\end{document}


\title{\textbf{Supplemental Material \\
			for \\
			``Intertwined Hyperferroelectricity, Tunable Multiple Topological Phases and Giant Rashba Effect in Wurtzite LiZnAs"} 
	} 
	

	
	\author{Saurav Patel}
	\affiliation{Department of Physics, Faculty of Science, The Maharaja Sayajirao University of Baroda, Vadodara-390002, Gujarat, India.}
	
	\author{Paras Patel}
	\affiliation{Department of Physics, Faculty of Science, The Maharaja Sayajirao University of Baroda, Vadodara-390002, Gujarat, India.}
	
	\author{Shaohui Qiu}
	\affiliation{Department of Physics, University of Arkansas, Fayetteville, Arkansas 72701, USA.}
	
	\author{Prafulla K. Jha}
	\email{prafullaj@yahoo.com}  
	\affiliation{Department of Physics, Faculty of Science, The Maharaja Sayajirao University of Baroda, Vadodara-390002, Gujarat, India.}

\maketitle
\clearpage

\textbf{Section I:} The phonon dispersion curves of \textit{P6$_3$mc} phase after structural phase transition at 8 GPa.

\begin{figure}[H] 
	\centering
	\includegraphics[width=0.50\linewidth]{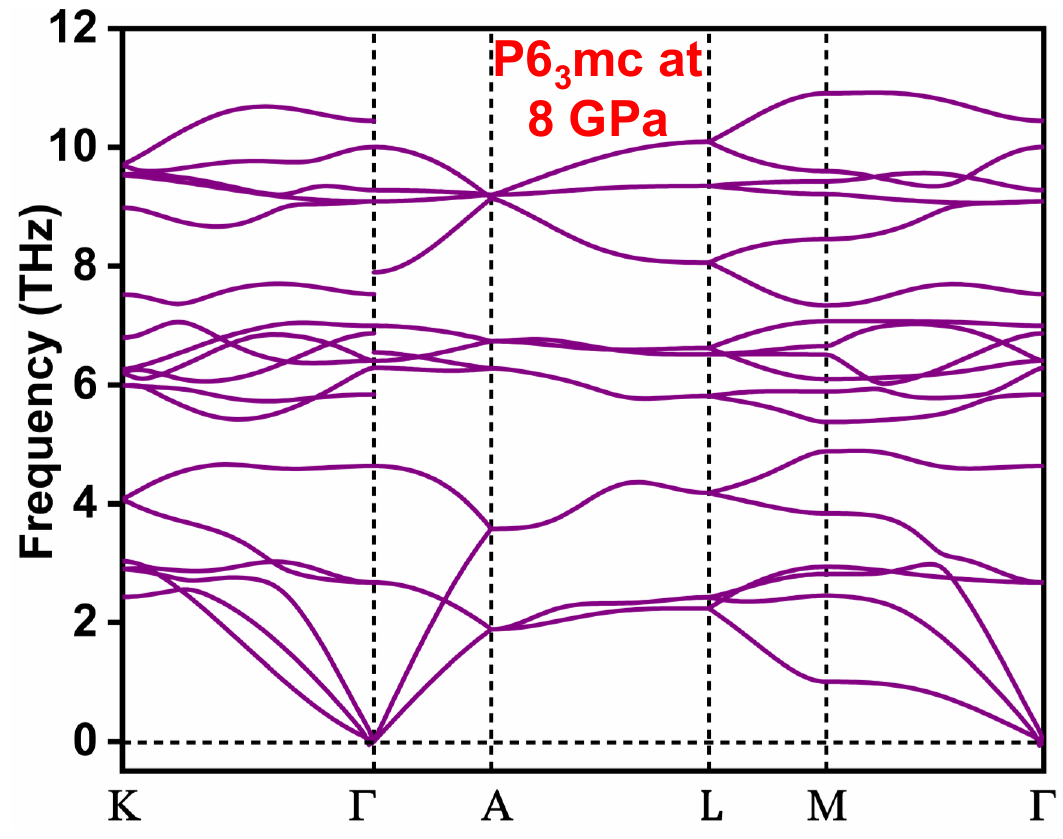}
	\caption{The calculated phonon dispersion curves of polar \textit{P6$_3$mc} phase at 8 GPa.}
	\label{Fig:phonon at 8 GPa}
\end{figure}

\textbf{Section II:} The calculated electronic band structure of \textit{P6$_3$mmc} and \textit{P6$_3$mc} phases. The band gap of 0.52 eV in \textit{P6$_3$mmc} and 0.34 eV in \textit{P6$_3$mc} indicates the insulating nature and thus facilitates the use of Berry-phase approach to determine the electronic polarization.

\begin{figure}[H] 
	\centering
	\includegraphics[width=1.00\linewidth]{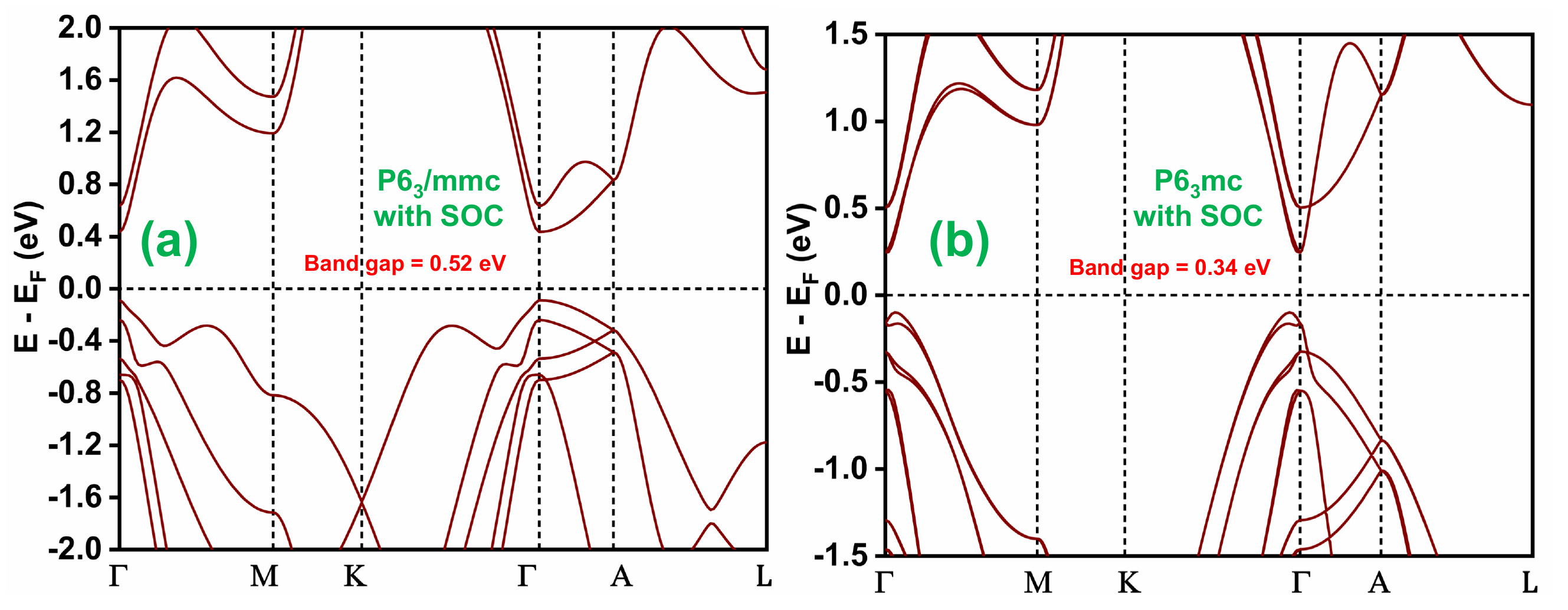}
	\caption{The relativistic band structure of non-polar \textit{P6$_3$/mmc} (a) and polar \textit{P6$_3$mc} (b) phases, respectively.}
	\label{Fig:full BZ electronic bands}
\end{figure}

\bibliography{ref}